\documentclass[aps,pre,showpacs,twocolumn]{revtex4}
\usepackage{bbm}
\usepackage{mathrsfs}
\usepackage{epsfig,psfrag}
\usepackage{amsmath,amsfonts,amssymb}
\usepackage[usenames]{color}

\def\be{\begin{equation}}
\def\ee{\end{equation}}
\def\bea{\begin{eqnarray}}
\def\eea{\end{eqnarray}}

\begin{document}

\preprint{draft}

\title{Conformal Invariance of Iso-height Lines in two-dimensional KPZ Surface }
\author{A. A. Saberi $^{1}$, M. D. Niry $^{1}$, S. M. Fazeli $^{1}$, M. R. Rahimi Tabar $^{1,2}$, and S. Rouhani $^{1}$}

\address{$^{1}$Department of Physics, Sharif University of Technology, Tehran
11155-9161, Iran \\ $^{2}$ Carl von Ossietzky University, Institute
of Physics, D-26111 Oldendurg, Germany }

\date{\today}

\begin{abstract}
The statistics of the iso-height lines in (2+1)-dimensional
Kardar-Parisi-Zhang (KPZ) model is shown to be conformal invariant
and equivalent to those of self-avoiding random walks. This leads to
a rich variety of new exact analytical results for the KPZ dynamics.
We present direct evidence that the iso-height lines can be
described by the family of conformal invariant curves called
Schramm-Loewner evolution (or $SLE_\kappa$) with diffusivity
$\kappa=8/3$. It is shown that the absence of the non-linear term in
the KPZ equation will change the diffusivity $\kappa$ from $8/3$ to
$4$, indicating that the iso-height lines of the Edwards-Wilkinson
(EW) surface are also conformally invariant, and belong to the
universality class of the domain walls in the O(2) spin model.
\end{abstract}


\maketitle

Recently, it was shown that the statistics of the zero-vorticity
lines in inverse cascade of two dimensional (2D) Navier-Stokes
turbulence is conformally invariant and belongs to the percolation
universality class \cite{bernard1}. The same issue has been studied
for zero-temperature isolines in the inverse cascade of surface
quasigeostrophic turbulence \cite{bernard2}, domain walls of spin
glasses \cite{spin glass} and the nodal lines of random wave
functions \cite{Keatin}. Moreover, it has been shown recently that
the statistics of the iso-height lines on the experimental $WO_3$
grown surface is the same as domain walls statistics in the critical
Ising model as well as those of Ballistic Deposition (BD) model \cite{WO3}.\\
Evidence of conformal invariance in the geometrical features of such
complex nonlinear systems have been provided, in the continuum
limit, by stochastic (Schramm) Loewner evolution, i.e.,
$SLE_{\kappa}$, where $\kappa$ is the diffusivity
\cite{schramm,cardy}. Schramm and Sheffield showed that the contour
lines in a two-dimensional discrete Gaussian free field are
statistically equivalent to $SLE_{4}$ \cite{Schramm-Sheffield}.
Moreover, it is shown that the restriction property only applies in
the case for $\kappa=8/3$ \cite{Schramm-Lawler}. Since self-avoiding
random walk (SAW) satisfies the restriction property, it is
conjectured in the scaling limit to fall in the SLE class with
$\kappa=8/3$ \cite{Tom-Kennedy}. The scaling limit of SAW in the
half-plane has been proven to exist \cite{G. Lawler} but there is no
general proof of its existence.

In this Letter we investigate numerically the iso-height lines of
the (2+1)-dimensional Kardar-Parisi-Zhang (KPZ) model \cite{Kardar},
and study their possible conformal invariance. It is shown that the
KPZ's iso-height lines are equivalent to self-avoiding walks, and
that the iso-height lines in the 2D-KPZ surface are $SLE_{8/3}$
curves. For the Edwards-Wilkinson (EW) interface (the KPZ model
without the nonlinear term) the iso-height lines fall in the
universality class of the interfaces in the $O(2)$ model, and can be
described by $SLE_4$.

The KPZ equation is given by \be\label{eqKPZ} \frac{\partial
h(\textbf{x},t)}{\partial t}=\nu\nabla^2h+\frac{\lambda}{2}
\mid\nabla h\mid^2+\eta(\textbf{x},t)\;. \ee The first term on the
r.h.s describes relaxation of the interface caused by a surface
tension $\nu$, and the nonlinear term is due to the lateral growth.
The noise $\eta$ is uncorrelated Gaussian white noise in both space
and time with zero average i.e., $
\langle\eta(\textbf{x},t)\rangle=0 $ and $\langle
\eta(\textbf{x},t)\eta(\textbf{x}',t')\rangle=2D\delta^d(\textbf{x}-\textbf{x}')\delta(t-t')$.
The KPZ equation is invariant under translations along both growth
direction and perpendicular to it, as well as time translation and
rotation \cite{stanley}. Rescaling the variables,
$h=\tilde{h}\sqrt{2D/\nu}$, and $t=\tilde{t}/\nu$, changes Eq.
(\ref{eqKPZ}) to, $\partial\tilde{h}(\textbf{x},\tilde{t})/
\partial{\tilde{t}}=\nabla^2\tilde{h}+\sqrt{\epsilon}\mid\nabla
\tilde{h}\mid^2+\tilde{\eta}(\textbf{x},\tilde{t})$, where
$\epsilon= \lambda^2D/2\nu^3$ and
$\langle\tilde{\eta}(\textbf{x},\tilde{t})
\tilde{\eta}(\textbf{x}',\tilde{t}')\rangle=\delta^d(\textbf{x}-\textbf{x}')
\delta(\tilde{t}-\tilde{t}')$. In the following, we work with the
single parameter $\epsilon$ and drop all the tildes for simplicity.

We have studied the rescaled KPZ equation on a square lattice with
periodic boundary conditions. The numerical integration was done
using the Runge-Kutta-Fehlberg scheme of orders $O(4)$ and $O(5)$
\cite{Toral}. This scheme controls automatically the integration
time step $\delta t$, such that the resulting height error $\delta
h$ (which is estimated by comparing the results obtained from the
$O(4)$ and $O(5)$ integrations) can be ignored at each time step. We
took the error to be less than 0.1, and we checked that smaller
values of $\delta h$ do not improve the precision of the computed
quantities. The noise $\eta$ was generated by the Box-Muller method.
To avoid the instabilities that may appear during the growth, we
used the algorithm introduced in \cite{Sarma}, where the term
$(1-c^{-1}e^{-cf})$, is used instead of the nonlinear term in Eq.
(\ref{eqKPZ}), i.e., $f=\mid\nabla h\mid^2$. Since $f\ll
w_L^2(\infty)$, where $ w_L(t)$ is the interface width of the system
with size  $L$ at time $t$, by keeping $c\ll w_L^{-2}(\infty)$, one
can control the possible numerical divergencies that may appear
during the integration. Clearly for very small $c$ this term
converges to $f$.\\
We have checked that the growth exponents for $(1+1)$-dimension are
obtained correctly (both roughness and growth exponents $\alpha=1/2$
and $\beta=1/3$ respectively), and the (2+1)-dimensional results are
given in Fig. 1, which are in good agreement with previous studies
\cite{Parisi}.

\begin{figure}\begin{center}
\includegraphics[width=3.4 in, height=3 in]{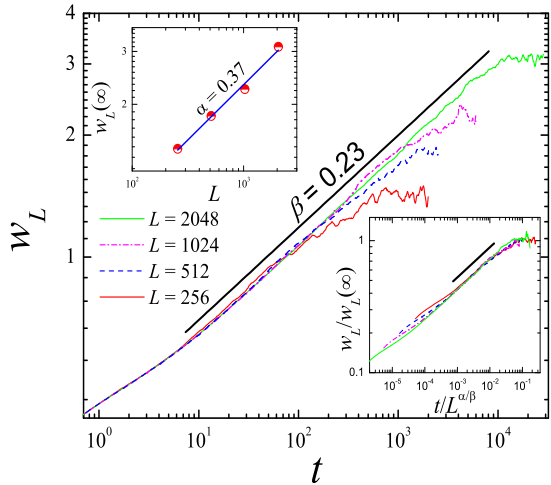}
\narrowtext\caption{\label{Fig1}(color online). Main frame:
Interface width $w_L(t)$ vs time $t$ of the KPZ equation in
(2+1)-dimensions and for $\epsilon=10$, for different square lattice
size $L$. The slope of the straight line yields the growth exponent
$\beta=0.23\pm0.01$. Upper-left inset: Saturation width
$w_L(\infty)$ for systems of different size $L$. The slope of the
solid-line fit yields the roughness exponent $\alpha=0.37\pm0.01$.
Lower-right inset: Rescaled $w_L$ vs rescaled $t$.}\end{center}
\end{figure}

We now consider the saturated 2D-KPZ surface and set its  mean
height to be zero, and attribute the same sign to the points that
have positive or negative heights. The same-sign regions (clusters)
and their boundaries (loops) were identified by the Hoshen-Kopelman
algorithm (Fig. 2).
\begin{figure}\begin{center}
\includegraphics[scale=0.65]{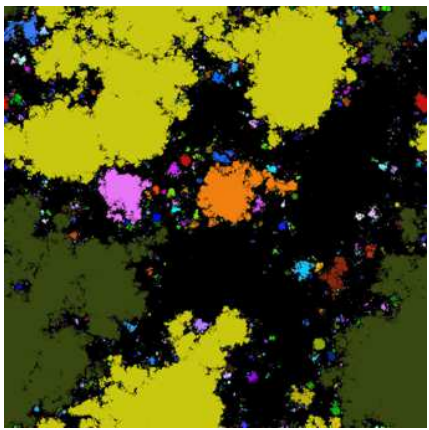}
\narrowtext\caption{\label{Fig2}(color online). The clusters with
positive heights are shown for the 2D-KPZ interface with different
colors. Negative-height regions are colored with black.}\end{center}
\end{figure}

\begin{figure}\begin{center}
\includegraphics[width=3.4 in, height=5.0 in]{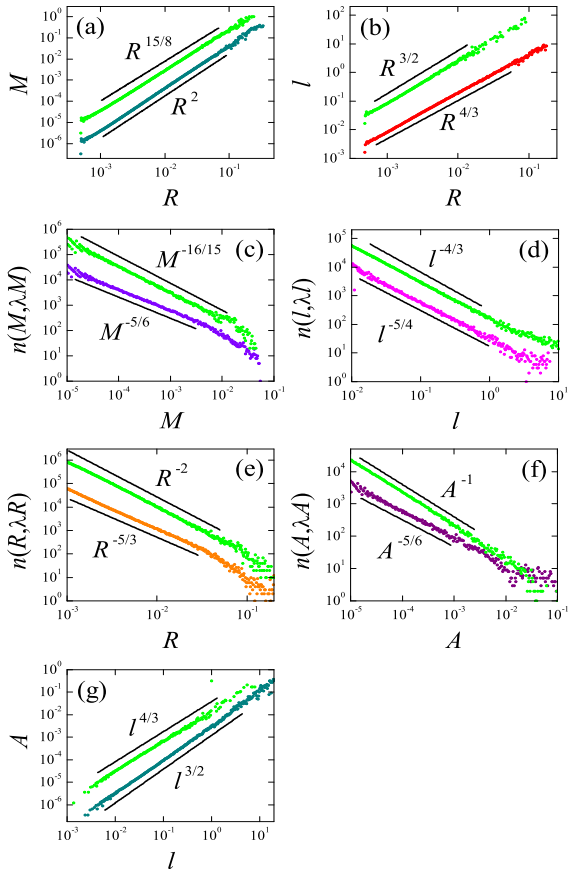}
\newline\narrowtext\caption{\label{Fig3}(color online).
Cluster and loop statistics for the iso-height lines of 2D-KPZ
surface and the EW surface are shown by the lower (different colors)
and upper (green) graphs, respectively. The results for the EW have
been shifted by a constant 1 in order to distinguish them from the
2D-KPZ results. (a) The average area $M$ vs the radius of gyration
$R$. (b) The length of a loop $l$ vs the radius of gyration $R$. (c)
Number of clusters of area between $M$ and $\lambda M$. (d) Number
of loops of length between $l$ and $\lambda l$. (e) Number of loops
of radius of gyration between $R$ and $\lambda R$. (f) Number of
loops of area between $A$ and $\lambda A$. (g) The average area of
loops $A$ vs the length $l$. In all figures $\lambda\simeq1.05$.
Solid lines show the corresponding analytical results for the SAWs
(bottom) and O(2) model (top). The error bars are almost the same
size as the symbols in the scaling regions, and have not been drawn.
The change in the exponents relative to the SAWs in Figs. (c), (d),
(e) and (f), is due to the roughness exponent
\cite{kondev}.}\end{center}
\end{figure}

To investigate the scaling behavior of such loop ensembles in the
2D-KPZ interface, a set of scaling exponents associated with cluster
and loop statistics were computed and checked, and are shown in Fig.
\ref{Fig3} \cite{kondev,Duplantier,Cardy2}. The estimated exponents
are in good agreement with the corresponding analytical results for
SAWs in the scaling limit. The fractal dimension $D_f$ of contour
lines obtained from the scaling relation between their length $l$
and radius of gyration $R$, i.e., $l\sim R^{D_f}$, is given by
$D_f=1.33\pm0.01$ which is in agreement with the one obtained by the
box-counting method for the largest contour lines, $D_f=1.33\pm
0.02$. Comparing with the known fractal dimension of $SLE_\kappa$
curves $D_f=1+\kappa/8$ for $0\leq\kappa\leq8$ , the contour lines
may have conformal invariant scaling limit according to $SLE_{8/3}$.

\begin{figure}\begin{center}
\includegraphics[width=3.4 in, height=2.8 in]{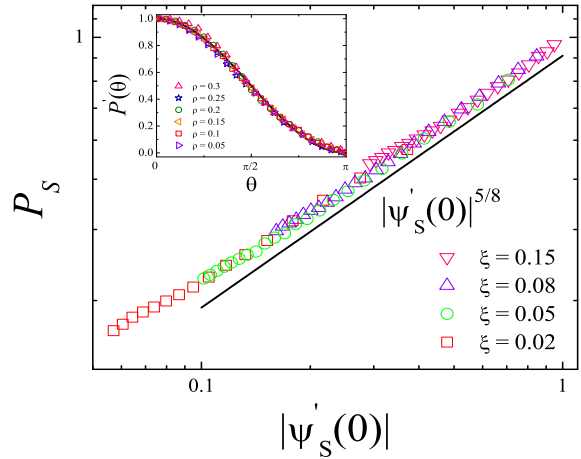}
\newline\narrowtext\caption{\label{Fig4}(color online). Main frame: The
probability that a contour line of 2D-KPZ surface in the upper
half-plane does not hit the slits of height $0.05\leq h\leq0.5$
placed at $\xi$ from origin on the real axis vs
$|\Psi'_S(0)|=|\xi|(\xi^2+h^2)^{-1/2}$. The solid line is the
corresponding analytical prediction for $SLE_{8/3}$ curves. The
error bars are almost the same size as the symbols in the scaling
region. Inset: The probability that such a contour line passes to
the left of a point $z=\rho e^{i\theta}$, for $\rho=0.05, 0.1, 0.15,
0.2, 0.25$ and $0.3$. The solid line shows the prediction of SLE for
$\kappa=8/3$.}\end{center}
\end{figure}

The quantity that can confirm the SAW property of the contour lines
is the restriction property. Suppose that $S$ is a hull in the
upper-half plane $\mathbb{H}$ which is bounded away from the origin,
and $\gamma$ is a simple SLE curve in $\mathbb{H}$ with $\kappa\leq
4$. Let $\Psi_S$ be a unique conformal map of $\mathbb{H}\backslash
S$ onto $\mathbb{H}$, such that $\Psi_S(0)=0$,
$\Psi_S(\infty)=\infty$ and $\Psi'_S(\infty)=1$. The restriction
property states that the distribution of curves conditioned not to
hit $S$ is the same as the distribution of curves in the domain
$\mathbb{H}\backslash S$. This happens only for $\kappa=8/3$, and it
is shown that \cite{Schramm-Lawler} the probability that a curve
does not hit the hull $S$ is
\be \label{Restriction} P[\gamma \cap
S=\emptyset]=|\Psi_S'(0)|^{5/8}.
\ee
To examine this property directly for contour lines on the saturated
2D-KPZ surface, we proceed as follows. First, we identify all the
cluster boundaries (contour lines): for each cluster an explorer
walks on the zero height line as keeping the sites with positive
height on the right. Then, we consider an arbitrarily placed
straight line for each curve as a real axis and cut the portion of
the curve above it. Using this procedure, we obtain an ensemble of
contour lines in the half-plane which start at the origin and end on
the real axis $x_\infty$. To obtain curves whose size is of order
one, we rescale them by a factor of $N^{\nu}$, where $\nu=1/D_f$.
Second, we consider the hull $S$ as a slit placed at various
distances $\xi$ from the origin and various heights $h$, for which
the map $\Psi_S$ is defined by,
$\Psi_S(z)=\xi+\sqrt{(z-\xi)^2+h^2}$. After mapping the curves by
$\varphi(z)=x_\infty z/(x_\infty-z)$ \footnote{We have used this map
to ensure that the curves begin at the origin and end at infinity,
the so-called chordal $SLE_\kappa$. Also, we have used these chordal
curves to measure the left-passage probability. To avoid numerical
errors, only the part of the curves corresponding to capacity $t\leq
0.3$ were used.}, we have checked Eq. (\ref{Restriction}) for the
contour lines of the 2D-KPZ surface. As shown in Fig. \ref{Fig4},
the result is consistent with Eq. (\ref{Restriction}) and implies
the connection between the contour lines and both the SAWs and
$SLE_{8/3}$.

Since the restriction property only holds for $SLE_{8/3}$ curves, we
test the probability that an SLE curve passes to the left of a given
point $z=\rho e^{i\theta}$, where $\theta$ is the angle between the
point and the origin, and $\rho$ is the distance from the origin
inside the upper half-plane. Given scale invariance, this
probability is independent of $\rho$, and the theory of SLE predicts
\cite{Left} that \be \label{Left}
P^\prime_{\kappa}(\theta)=\frac{1}{2}+\frac{\Gamma\left(\frac{4}{\kappa}\right)}{\sqrt{\pi}\Gamma\left(\frac{8-\kappa}{2\kappa}\right)}
{}_2F_1\left(\frac{1}{2},\frac{4}{\kappa};\frac{3}{2};-cot^2(\theta)\right)cot(\theta).
\ee Here, ${}_2F_1$ is the hypergeometric function. The computed
$P^\prime_{\kappa}(\theta)$ for the contour lines of the 2D-KPZ
surface is also consistent with the analytical form with
$\kappa=8/3\pm1/10$ (Fig. \ref{Fig4}).

These results strongly suggest that the iso-height lines might be,
in the scaling limit, conformally invariant, giving rise to the SLE
curves with $\kappa=8/3$. To examine this suggestion directly, we
can extract the Loewner driving function $\zeta$ of the curves using
the successive conformal maps. We use the algorithm introduced by
Bernard \textit{et al.} \cite{bernard2} based on the approximation
that driving function is a piecewise constant function. Each curve
is parameterized by a dimensionless parameter $t$, to be
distinguished from time in (\ref{eqKPZ}). The procedure is based on
applying the map $G_{t,\zeta}=x_\infty \{\eta
x_\infty(x_\infty-z)+[x_\infty^4(z-\eta)^2+4t(x_\infty-z)^2(x_\infty-\eta)^2]^{1/2}\}
/\{x_\infty^2(x_\infty-z)+[x_\infty^4(z-\eta)^2+4t(x_\infty-z)^2(x_\infty-\eta)^2]^{1/2}\}$
on all the points $z$ of the curve approximated by a sequence of
$\{z_0=0, z_1, \cdot\cdot\cdot, z_N=x_\infty\}$ in the complex
plane, where $\eta=\varphi^{-1}(\zeta)$ and again
$\varphi(z)=x_\infty z/(x_\infty-z)$. At each step, by using the
parameters $\eta_0=\varphi^{-1}(\zeta_0)=[Rez_1
x_\infty-(Rez_1)^2-(Imz_1)^2]/(x_\infty-Rez_1)$ and
$t_1=(Imz_1)^2x_\infty^4/\{4[(Rez_1-x_\infty)^2+(Imz_1)^2]^2\}$, one
point of the curve $z_0$ is swallowed and the resulting curve is
rearranged by one element shorter. This operation yields a set
containing $N$ numbers of $\zeta_k(t_k)$ for each curve.
\begin{figure}\begin{center}
\includegraphics[width=3.4 in, height=2.8 in]{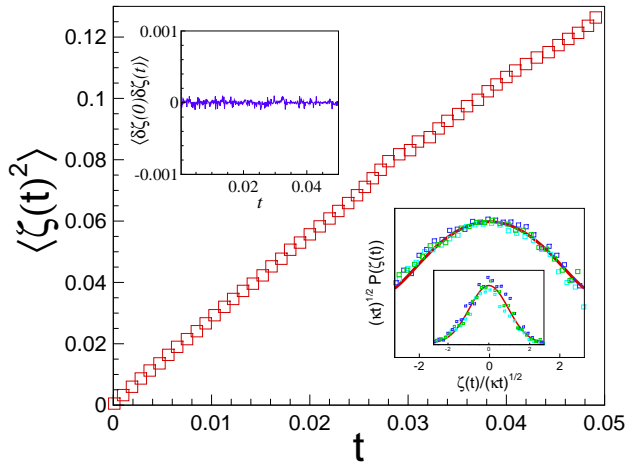}
\narrowtext\caption{\label{Fig5}(color online). Statistics of the
driving function $\zeta(t)$. Main frame: the linear behaviour of
$\langle\zeta(t)^2\rangle$ with the slope $\kappa=8/3\pm1/10$.
Lower-right inset: The probability density of $\zeta(t)$ rescaled by
its variance $\kappa$ at times $t=0.01, 0.015, 0.02$ which is
Gaussian. Upper-left inset: The correlation function of the
increments of the driving function $\delta\zeta(t)$.
Kolmogorov-Smirnov (K-S) goodness of fit for normal distribution of
the noise $\zeta(t)/\sqrt{\kappa t}$ for $\kappa=8/3$ and $0\leq
t\leq 0.05$ yields $K-S=0.017$}.\end{center}
\end{figure}
The next step is analyzing the ensemble of the driving functions
$\zeta(t)$ which can indicate, within the statistical errors,
whether the curves are SLE or not. As shown in Fig. \ref{Fig5}, the
statistics of the ensemble of $\zeta(t)$ converges to a Gaussian
process with variance $\langle\zeta^2(t)\rangle=\kappa t$ and
$\kappa=2.6\pm0.1$. This evidence certifies that the iso-height
lines of 2D-KPZ interface in the saturation regime appear to be
conformally invariant and are described by the $SLE_{8/3}$. The
above results were obtained for $\epsilon=10$; however, the same
analysis for growth surfaces with other values of $\epsilon$ (which
were checked for $\epsilon=5$ and $25$) indicates no changes.

\begin{figure}[h]\begin{center}
\includegraphics[scale=0.65]{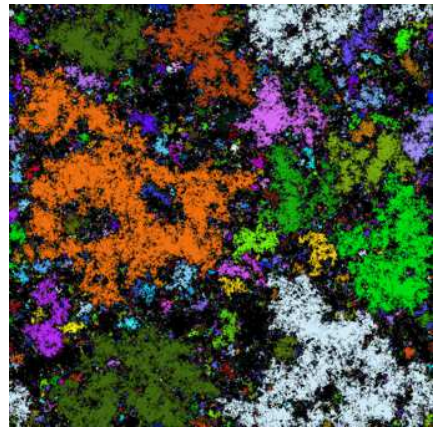}
\narrowtext\caption{\label{Fig6}(color online). Positive height
connected domains of 2D-KPZ interface without the non-linear term,
corresponding to the EW model. Negative height regions are
black.}\end{center}
\end{figure}

In the case of $\epsilon=0$, which corresponds to the EW model,
comparing Fig. \ref{Fig6} and Fig. \ref{Fig2}, indicates more
"porosity" in the clusters, which is indicative of changes in the
cluster boundaries' shape. As presented in Fig. \ref{Fig3}, the
cluster and loop statistics in this case are most consistent with
those for the $O(2)$ model. These lead to the conclusion that if one
assumes that the scaling limit of such contour lines exists, it
should belong to the $SLE_4$ curves.
We also checked this directly as above and found that the driving
function has Gaussian statistics with variance $\kappa=3.7\pm0.2$.

The height clusters and their boundary statistics, in the manner
presented here, can be applied to model the experimental grown
surfaces by using an ensemble of the grown samples in the saturated
regime. Since this analysis is far more accurate, it may be also
used to investigate wether a model belongs to a universality class
or not. For example, the small difference between the cluster
analysis of the BD model \cite{WO3, krug1, krug2} and KPZ equation
in two dimensions can be revealed.

We wish to thank J. Cardy for useful hints on this work. We also
thank S. Moghimi-Araghi, M. A. Rajabpour and M. Sahimi for useful
discussions.


\begin{thebibliography}{}
\bibitem{bernard1}   D. Bernard, \textit{et al.}, Nature Phys. \textbf{2}, 124
(2006).
\bibitem{bernard2}   D. Bernard, \textit{et al.}, Phys. Rev. Lett. \textbf{98}, 024501
(2007).
\bibitem{spin glass}  C. Amoruso, \textit{et al.}, Phys. Rev. Lett. \textbf{97}, 267202 (2006); D. Bernard,
\textit{et.al}, Phys. Rev. B \textbf{76}, 020403(R) (2007).
\bibitem{Keatin}  J. P. Keating, \textit{et al.},
Phys. Rev. Lett. \textbf{97}, 034101 (2006); E. Bogomolny,
\textit{et al.}, [nlin/0609017].
\bibitem{WO3} A. A. Saberi, \textit{et.al}, Phys. Rev. Lett, \textbf{100}, 044504 (2008).
\bibitem{schramm}  O. Schramm, Israel. J. Math. \textbf{128}, 221 (2000).
\bibitem{cardy}    J. Cardy, Ann. Physics \textbf{318}, 81 (2005); M. Bauer
and D. Bernard, Phys. Rep. \textbf{432}, 115 (2006); I. A. Gruzberg,
J. Phys. A: Math. Gen. \textbf{39}, 12601 (2006); H. C. Fogedby,
[cond-math/0706.1177].
\bibitem{Schramm-Sheffield}  O. Schramm and S. Sheffield,
[Math.PR/0605337] (2006).
\bibitem{Schramm-Lawler}  GF. Lawler, \textit{et al.}, J. Amer. Math. Soc. \textbf{16},
917 (2003).
\bibitem{Tom-Kennedy} T. Kennedy, Phys. Rev. Lett. \textbf{88}, 130601
(2002).
\bibitem{G. Lawler} G. Lawler, \textit{et al.}, arXiv:math/0204277.
\bibitem{Kardar} M. Kardar, G. Parisi and Y.-C. Zhang, Phys. Rev. Lett.
\textbf{56}, 889-892 (1986).
\bibitem{stanley} A. L. Barabási and H. E. Stanley, Fractal Concepts in
Surface Growth (Cambridge University Press, Cambridge, 1995).
\bibitem{Toral} M. S. Miguel and R. Toral, \textit{Instabilities and Nonequilibrium
Structures}, (Kluver Academic Publishers, vol. VI, p. 35, 2000).
\bibitem{Sarma} C. Dasgupta, \textit{et al.}, Phys. Rev. E
\textbf{55}, 2235 (1997); C. Dasgupta, \textit{et al.}, Phys. Rev. E
\textbf{54}, R4552 (1996).
\bibitem{Parisi} J. G. Amar, F. Family, Phys. Rev. A \textbf{41}, 6,
3399 (1990); E. Marinari, \textit{et al.}, J. Phys. A: Math. Gen.
\textbf{33}, 8181 (2000).
\bibitem{kondev} J. Kondev, C. L. Henley, Phys. Rev. Lett, \textbf{74}, 23,
4580 (1995).
\bibitem{Duplantier} B. Duplantier, Phys. Rev. Lett, \textbf{64}, 4,
493 (1990).
\bibitem{Cardy2} J. Cardy, Phys. Rev. Lett, \textbf{72}, 1580-1583 (1994).
\bibitem{Left} O. Schramm, Electron. Commun. Probab. \textbf{6}, 115 (2001).
\bibitem{krug1} P. Meakin and J. Krug, Europhys. Lett. \textbf{11}, 7-12 (1990).
\bibitem{krug2} P. Meakin and J. Krug, Phys. Rev. A \textbf{46}, 3390-3399 (1992).
\end{thebibliography}
\end{document}